\documentclass[fleqn,twoside]{article}
\usepackage{espcrc2}

\usepackage{graphicx}
\usepackage[figuresright]{rotating}

\newcommand{\AmS}{{\protect\the\textfont2
  A\kern-.1667em\lower.5ex\hbox{M}\kern-.125emS}}

\title{Five-Dimensional Cosmological Theory of Unified  
Space, Time and Velocity}

\author{Moshe Carmeli\address[MCSD]{Department of Physics, 
Ben Gurion University, \\
Beer Sheva 84105, Israel}}%
\begin{document}

\begin{abstract}
A five-dimensional cosmological theory of gravitation that unifies space, time
and velocity is presented. 
\vspace{1pc}
\end{abstract}

\maketitle

\section{INTRODUCTION}
In this lecture we introduce a five-dimensional cosmological theory of space, 
time and velocity. The added extra dimension of velocity to the usual 
four-dimensional spacetime will be evident in the sequel. Before introducing the
theory we have to deal, as usual, with coordinate systems in cosmology. Other
important basic issues will be dealt with later on.  

We will use {\it cosmic coordinate systems} that fill up spacetime. Given one system $x$, 
there is another one $x'$ that differs from the original one by a {\it Hubble
transformation} 
$x'=x+t_1v,\hspace{5mm} t_1=\mbox{\rm constant},$
where $v$ is a velocity parameter, and $y$ and $z$ are kept unchanged. A third system will be given by another
Hubble transformation,
$x''=x'+t_2v=x+(t_1+t_2)v.$ 

The cosmic coordinate systems are similar to the inertial coordinate systems,
but now the velocity parameter takes over the time parameter.
The analogous Galileo transformation to the latter that relates inertial 
coordinate systems is given, 
as is known, by 
$x''=x'+v_2t=x+(v_1+v_2)t.$

The universe expansion is also given by a formula of the above kind:
$x'=x+\tau v,$
where $\tau=H_0^{-1}$ in the limit of zero distance. However, the universe 
expansion is apparently incompatible with the Hubble spacetime transformation, 
namely one cannot add them. Thus, if we have 
$x''=x'+tv, \hspace{5mm} x'=\tau v,$
then 
$x''\neq\left(\tau+t\right)v.$
Rather, it is always
$x''=\tau v.$

This situation is like that we have with the propagation of light, 
$x''\neq\left(c+v\right)t,$
but it is always
$x''=ct$
in all inertial coordinate systems, and where $c$ is the speed of light in
vacuum.

The constancy of the speed of light and the validity of the laws of nature in
inertial coordinate systems, though they are both experimentally valid, they
are not compatible with each other. We have the same situation in cosmology; 
the constancy of the Hubble constant in the zero-distance limit, and the
validity of the laws of nature in cosmic coordinate systems, though both are
valid, they are incompatible with each other. 

In the case of light propagation, one has to abandon the Galileo 
transformation in favor of the Lorentz transformation. In cosmology one has
to give up the Hubble transformation for the cosmological  
transformation given by \cite{1}
\begin{equation}
\begin{array}{l}
x'=\displaystyle\normalsize\frac{x-tv}{\sqrt{1-t^2/\tau^2}},\hspace{5mm} v'=
\displaystyle\normalsize\frac{v-tx/\tau^2}{\sqrt{1-t^2/\tau^2}},\\ 
y'=y, \hspace{5mm} z'=z,\\\end{array}
\end{equation}
for the case with fixed $y$ and $z$. 

As is well known, the flat spacetime line element in special relativity is 
given by 
$ds^2=c^2dt^2-(dx^2+dy^2+dz^2).$
The cosmological flat spacetime line element is given, accordingly, by
\begin{equation}
ds^2=\tau^2dv^2-(dx^2+dy^2+dz^2).
\end{equation}

The special-relativistic line element is invariant under the Lorentz 
transformation. So is the cosmological line element: it is invariant under the
Lorentz-like cosmological transformation. The first keeps invariant the propagation 
of light, whereas the second keeps invariant the expansion of the universe.
At small velocities with respect to the speed of light, $v\ll c$, the Lorentz
transformation goes over to the nonrelativistic Galileo transformation. So is
the situation in cosmology: the Lorentz-like cosmological transformation goes over 
to the nonrelativistic Hubble transformation that is valid for cosmic times
much smaller than the Hubble time, $t\ll\tau$.
\section{Universe with Gravitation}
The universe is, of course, not flat but filled up with gravity. When 
gravitation is invoked, the above spaces become curved Riemanian with the 
line element 
$ds^2=g_{\mu\nu}dx^\mu dx^\nu,$
where $\mu$, $\nu$ take the values 0, 1, 2, 3, 4. The coordinates are: $x^0=
ct$, $x^1,x^2,x^3$ are spatial coordinates and $x^4=\tau v$. The 
signature is
$(+---+)$. The metric tensor $g_{\mu\nu}$ is symmetric and thus we have 
fifteen independent components. They will be a solution of the Einstein field 
equations in five dimensions.

The
five-dimensional field equations will not explicitely include a cosmological
constant, the latter is derivable from the theory. Our cosmological
constant will be equal to $\Lambda=3/\tau^2\approx 1.934\times 
10^{-35}$s$^{-2}$ (for $H_0=70$km/s-Mpc). 
This should
be compared with results of the experiments recently done with the supernovae 
which suggest
the value of $\Lambda\approx 10^{-35}$s$^{-2}$. Our cosmological constant
is derived from the theory itself which is part of the classification of
the cosmological spaces to describe deccelerating, constant or accelerating
universe. We now discuss some basic questions that are encountered in going
from four to five dimensions.

First we have to iterate what do we mean by coordinates in general and how
one measures them. The time coordinate is measured by clocks as was emphasized
by Einstein repeatedly \cite{2,3}. So are the spatial coordinates: they are 
measured by meters,
as was originally done in special relativity theory by Einstein, or by use of
Bondi's more modern version of k-calculus \cite{4,5}. 

But how about the velocity as an
independent coordinate? One might incline to think that if we know the
spatial coordinates then the velocities are just the time-derivative of the
coordinates and they are not independent coordinates. This is, indeed, the
situation for a dynamical system when the coordinates are given as functions 
of the time. But in general the situation is different, especially in 
cosmology. Take, for instance, the Hubble law $v=H_0x$. Obviously $v$ and $x$
are independent parameters and $v$ is not the time derivative of $x$. Basically
one can measure $v$ by instruments like those used by traffic police. 

To finish this section we discuss the important concept of the energy density
in cosmology. We use the Einstein field equations, in which the right-hand 
side includes the energy-momentum tensor. For fields other than gravitation,
like the electromagnetic field, this is a straightforward expression that
comes out as a generalization to curved spacetime of the same tensor 
appearing in special-relativistic electrodynamics. However, when dealing with
matter one should construct the energy-momentum tensor according to the 
physical situation (see, for example, Fock, Ref. 16). Often a special 
expression for the
mass density $\rho$ is taken for the right-hand side of Einstein's equations, 
which sometimes is expressed as a $\delta$-function. 

In cosmology we also have the situation where the mass density is put on the
right-hand side of the Einstein field equations. There is also the critical
mass density $\rho_c=3/8\pi G\tau^2$, the value of which is about $10^{-29}$
g/cm$^3$, just a few hydrogen atoms per cubic meter throughout the cosmos. If
the universe average mass density $\rho$ is equal to $\rho_c$ then the 
universe will have a constant expansion. A
deviation from this necessiates an increase or decrease 
from $\rho_c$. That is to say that 
$\rho_{eff}=\rho-\rho_c$ 
is the active or the effective 
mass density that causes the universe not to have a constant expansion. Accordingly,
one should use $\rho_{eff}$ in the right-hand side of the Einstein  field 
equations. Indeed, we will use such a convention throughout this paper. The
subtraction of $\rho_c$ from $\rho$ in not significant for celestial bodies
and makes no difference.  
\section{The Accelerating Universe}
In the last two sections we gave arguments to the fact that the universe 
should be presented in five dimensions, even though the standard cosmological
theory is obtained from Einstein's four-dimensional general relativity theory.
The situation here is similar to that prevailed before the advent of ordinary
special relativity. At that time the equations of electrodynamics, written in
three dimensions, were well 
known to predict that the speed of light was constant. But that was not the 
end of the road. The abandon of the concept of absolute space along with the
constancy of the speed of light led to the four-dimensional notion. In cosmology
now, we have to give up the notion of absolute cosmic time. Then this with the
constancy of the Hubble constant in the limit of zero distance leads us to a
five-dimensional presentation of cosmology.  

We recall that the field equations are those of Einstein in five dimensions,
$R_\mu^\nu-\frac{1}{2}\delta_\mu^\nu R=\kappa T_\mu^\nu,$
where Greek letters $\alpha,\beta,\cdots,\mu,\nu,\cdots=0,1,2,3,4$. The 
coordinates are $x^0=ct$, $x^1$, $x^2$ and $x^3$ are space-like coordinates,
$r^2=(x^1)^2+(x^2)^2+(x^3)^2$, and $x^4=\tau v$. The metric used is
given by $g_{00}=1+\phi$, $g_{kl}=-\delta_{kl}$, $g_{44}=1+\psi$, other 
components are zero.
We will keep only linear terms. The 
components of the Ricci tensor and the Ricci scalar are given by
\begin{equation}
\begin{array}{l}
R_0^0=\displaystyle\normalsize\frac{1}{2}\left(\nabla^2\phi-\phi_{,44}-\psi_{,00}\right),\\
R_0^n=\displaystyle\normalsize\frac{1}{2}\psi_{,0n},\hspace{2mm} 
R_n^0=-\displaystyle\normalsize\frac{1}{2}\psi_{,0n},
\hspace{2mm} R_0^4=R_4^0=0,\\
R_m^n=\displaystyle\normalsize\frac{1}{2}\left(\phi_{,mn}+\psi_{,mn}\right),\\
R_n^4=-\displaystyle\normalsize\frac{1}{2}\phi_{,n4},\hspace{2mm} 
R_4^n=\displaystyle\normalsize\frac{1}{2}\phi_{,n4}.\\
$$R_4^4=\displaystyle\normalsize\frac{1}{2}\left(\nabla^2\psi-\phi_{,44}-\psi_{,00}\right),\\
\end{array}
\end{equation}
\begin{equation}
R=\nabla^2\phi+\nabla^2\psi-\phi_{,44}-\psi_{,00}.
\end{equation}
In the above equations $\nabla^2$ is the ordinary three-dimensional Laplace 
operator. 

The line element in five dimensions is given by 
\begin{equation}
ds^2=(1+\phi)dt^2-dr^2+(1+\psi)dv^2,
\end{equation}
where $dr^2=(dx^1)^2+(dx^2)^2+(dx^3)^2$, and where $c$ and $\tau$ were taken,
for brevity, as equal to 1. The 
line element (5) represents a spherically symmetric universe.

The expansion of the universe (the Hubble expansion) is recorded at a definite 
fixed time and thus $dt=0$. Accordingly, taking into account $d\theta=d\phi=0$, 
Eq. (5) gives the following equation
for the expansion of the universe at a certain moment,
\begin{equation}
-dr^2+(1+\psi)dv^2=0,
\end{equation} 
and thus 
\begin{equation}
\left(dr/dv\right)^2=1+\psi.
\end{equation}
To find $\psi$ we solve the Einstein field equation (noting that $T_0^0=
g_{0\alpha}T^{\alpha 0}\approx T^{00}=\rho(dx^0/ds)^2\approx c^2\rho$, or 
$T^0_0\approx\rho$ in units with $c=1$):  
\begin{equation}
R_0^0-\frac{1}{2}\delta_0^0R=8\pi G\rho_{eff}=8\pi G\left(\rho-\rho_c\right),
\end{equation}
where $\rho_c=3/8\pi G\tau^2$. 

A simple calculation using Eqs. (3) and (4) then yields 
\begin{equation}
\nabla^2\psi=6(1-\Omega), 
\end{equation}
where $\Omega=\rho/\rho_c$.

The solution of the field equation (9) is given by 
\begin{equation}
\psi=(1-\Omega)r^2+\psi_0, 
\end{equation}
where the first part on the right-hand side is a solution for the 
non-homogeneous Eq. (9), and $\psi_0$ represents a solution to its 
homogeneous part, i.e. $\nabla^2\psi_0=0$. A solution for $\psi_0$ can be
obtained as an infinite series in powers of $r$. The only term that is left is
of the form $\psi_0=-K_2/r$, where $K_2$ is a constant whose value can easily 
be shown to be the Schwartzschild radius, $K_2=2GM$. We therefore have
\begin{equation}
\psi=(1-\Omega)r^2-2GM/r.
\end{equation} 

The universe expansion is therefore given by 
\begin{equation}
\left(dr/dv\right)^2=1+\left(1-\Omega\right)r^2-\frac{2GM}{r}.
\end{equation}
For large $r$ the last term on the right-hand side of (12) can be neglected,
and therefore 
\begin{equation}
\left(dr/dv\right)^2=1+(1-\Omega)r^2,
\end{equation}
or 
\begin{equation}
dr/dv=\left[1+\left(1-\Omega\right)r^2\right]^{1/2}.
\end{equation}
Inserting now the constants $c$ and $\tau$ we finally obtain for the expansion
of the universe
\begin{equation}
dr/dv=\tau\left[1+\left(1-\Omega\right)r^2/c^2\tau^2\right]^{1/2}.
\end{equation}
This result is exactly that obtained by Behar and Carmeli (BC) (Eq. 5.10) when 
the non-relativistic  relation $z=v/c$, where $z$ is the redshift parameter, 
is inserted in the previous result \cite{6}.

The second term in the square bracket of (15) represents the deviation
from constant expansion due to gravity. For without this term, Eq. (15)
reduces to $dr/dv=\tau$, thus $r=\tau v+const$. The constant can be taken zero
if one assumes, as usual, that at $r=0$ the velocity should also vanish. 
Accordingly we have $r=\tau v$ or $v=\tau^{-1}r$. Hence when 
$\Omega=1$, namely when $\rho=\rho_c$, we have a constant expansion.

The equation of motion (15) can be integrated exactly. The
results are: \newline
For the $\Omega>1$ case 
\begin{equation}
r(v)=\left(c\tau/\alpha\right)\sin\left(\alpha v/c\right); \hspace{2mm}
\alpha=\left(\Omega-1\right)^{1/2}.
\end{equation}
This is obviously a decelerating expansion.\newline
For $\Omega<1$,
\begin{equation}
r(v)=\left(c\tau/\beta\right)\sinh\left(\beta v/c\right); \hspace{2mm}
\beta=\left(1-\Omega\right)^{1/2}.
\end{equation}
This is now an accelerating expansion. 

For $\Omega=1$ we have, from Eq. (15),
\begin{equation}
d^2r/dv^2=0,
\end{equation}
whose solution is, of course,
\begin{equation}
r(v)=\tau v,
\end{equation}
and this is a constant expansion. It will be noted that the last solution
can also be obtained directly from the previous two cases for $\Omega>1$ and
$\Omega<1$ by going to the limit $v\rightarrow 0$, using L'Hospital's lemma,
showing that our solutions are consistent. 

It has been shown in BC that the 
constant expansion is just a transition stage between the decelerating and the
accelerating expansions as the universe evolves toward its present situation.
This occured at 8.5 Gyr ago at a time the cosmic radiation
temperature was 143K \cite{6}. 

In order to decide which of the three cases is the appropriate one at the present
time, it will be convenient to write the solutions (16), (17) and (19) in the ordinary
Hubble law form $v=H_0r$. Expanding Eqs. (16) and (17) and keeping the
appropriate terms then yields 
\begin{equation}
r=\tau v\left(1-\alpha^2v^2/6c^2\right),
\end{equation}
\begin{equation}
r=\tau v\left(1+\beta^2v^2/6c^2\right),
\end{equation}
for the $\Omega>1$ and $\Omega<1$ cases, respectively.
Using now the expressions for $\alpha$ and $\beta$ in Eqs. (20) and (21),
then both of the latter can be reduced into the single equation
\begin{equation}
r=\tau v\left[1+\left(1-\Omega\right)v^2/6c^2\right].
\end{equation}
Inverting now this equation by writing it in the form $v=H_0r$, we obtain in
the lowest approximation for $H_0$ 
\begin{equation}
H_0=h\left[1-\left(1-\Omega\right)v^2/6c^2\right],
\end{equation}
where $h=1/\tau$. Using $v\approx r/\tau$, or $z\approx v/c$, we also obtain
\begin{equation}
\begin{array}{l}
H_0=h\left[1-\left(1-\Omega\right)r^2/6c^2\tau^2\right]\\
\hspace{15mm}=h\left[1-\left(1-\Omega\right)z^2/6\right].\\\end{array}
\end{equation}
 
As is seen $H_0$ depends on the distance, or equivalently, on the redshift.
Cosequently $H_0$ has meaning only in the limits $r\rightarrow 0$ and $z
\rightarrow 0$, namely when measured {\it locally}, in which case it becomes 
the constant $h$. This is similar to the situation with respect to the speed 
of light when measured globally in the presence of gravitational field as the 
ratio between distance and time, the result usually depends on these 
parameters. Only in the limit one obtains the constant speed of light in 
vacuum ($c\approx 3\times 10^{10}$cm/s). 

Accordingly, $H_0$ is intimately related to the sign
of the factor $(1-\Omega)$. If measurements of $H_0$ indicate that it 
increases with the redshift parameter $z$ then the sign of $(1-\Omega)$ is 
negative, namely $\Omega>1$. If, however, $H_0$ decreases when $z$ increases
then the sign of $(1-\Omega)$ is positive, i.e. $\Omega<1$. The possibility of
$H_0$ not to depend on the redshift parameter indicates that $\Omega=1$. In
recent years different measurements were obtained for $H_0$ with the 
so-called 
``short" and ``long" distance scales, in which higher values of $H_0$ were 
obtained for the short distances and the lower values for $H_0$ corresponded 
to the long distances \cite{7,8,11,12,13,14,15,16}. Indications are that 
the longer the distance of 
measurement,
the smaller the value of $H_0$. If one takes these experimental results 
seriously, then that is possible only for the case in
which $\Omega<1$, namely when the universe is at an accelerating expansion 
phase, and the universe is thus open. 
\section{The Cosmological Constant}
First, a historical remark. In order to allow the existence of a static
solution for  the gravitational field equations, Einstein made a modification
to his original equations by adding a cosmological term, 
\begin{equation}
R_{\mu\nu}-\frac{1}{2}g_{\mu\nu}R+\Lambda g_{\mu\nu}=\kappa T_{\mu\nu},
\end{equation}
where $\Lambda$ is the cosmological constant and $\kappa=8\pi G$ ($c$ is taken 
as 1). For a homogeneous and 
isotropic universe with the line element \cite{19,20}
\begin{equation}
\begin{array}{l}
ds^2=dt^2\\
-a^2\left(t\right)R_0^2\left[\frac{dr^2}{1-kr^2}+r^2\left(
d\theta^2+\sin^2\theta d\phi^2\right)\right],\\\end{array}
\end{equation}
where $k$ is the curvature parameter ($k=1,0,-1$) and $a(t)=R(t)/R_0$ is the 
scale factor, with the energy-momentum tensor 
\begin{equation}
T_{\mu\nu}=\left(\rho+p\right)u_\mu u_\nu+pg_{\mu\nu},
\end{equation}
Einstein's equations (5.1) reduce to the two Friedmann equations
\begin{equation}
H^2\equiv\left(\frac{\dot{a}}{a}\right)^2=\frac{\kappa}{3}\rho+
\frac{\Lambda}{3}-\frac{k}{a^2R_0^2},
\end{equation}
\begin{equation}
\frac{\ddot{a}}{a}=-\frac{\kappa}{6}\left(\rho+3p\right)+\frac{\Lambda}{3}.
\end{equation}
These equations admit a static solution $(\dot{a}=0)$ with $k>0$ and 
$\Lambda>0$. After Hubble's discovery that the universe is expanding, the role
of the cosmological constant to allow static homogeneous solutions to Einstein's
equations in the presence of matter, seemed to be unnecessary. For a long
time the cosmological term was considered to be of no physical interest in 
cosmological problems.

From the Friedmann equation (28), for any value of the Hubble parameter $H$
there is a critical value of the mass density such that the spatial geometry
is flat ($k=0$), $\rho_c=3H_0^2/\kappa$. One usually 
measures the total
mass density in terms of the critical density $\rho_c$ by means of the 
density parameter $\Omega=\rho/\rho_c$. 

In general, the mass density $\rho$ includes contributions from various
distinct components. From the point of view of cosmology, the relevant aspect 
of each component is how its energy density evolves as the universe expands.
In general, a positive $\Lambda$ causes acceleration to the universe expansion,
whereas a negative $\Lambda$ and ordinary matter tend to decelerate it. 
Moreover, the relative contributions of the components to the energy density
change with time. For $\Omega_\Lambda<0$, the universe will always recollapse
to a Big Crunch. For $\Omega_\Lambda>0$ the universe will expand forever
unless there is sufficient matter to cause recollapse before $\Omega_\Lambda$
becomes dynamically important. For $\Omega_\Lambda=0$ we have the familiar
situation in which $\Omega_M\leq 1$ universes expand forever and $\Omega_M>1$
universes recollapse. (For more details see the paper by Behar and Carmeli,
Ref. 6.)

Recently two groups, the {\it Supernova Cosmology Project Collaboration} and 
the {\it High-Z Supernova Team Collaboration}, presented evidence that the 
expansion of the universe is accelerating \cite{21,22,23,24,25,26,27}.
These teams have measured the distances to cosmological supernovae by using 
the fact that the intrinsic luminosity of Type Ia supernovae is closely 
correlated with their decline rate from maximum brightness, which can be
independently measured. These measurements, combined with redshift data for
the supernovae, led to the prediction of an accelerating universe. Both
teams obtained 
\begin{equation}
\Omega_M\approx 0.3,\hspace{5mm}\Omega_\Lambda\approx 0.7,
\end{equation}
and strongly ruled out the traditional ($\Omega_M$, $\Omega_\Lambda$)=(1, 0)
universe. This value of the density parameter $\Omega_\Lambda$ corresponds to
a cosmological constant that is small but nonzero and positive,
\begin{equation}
\Lambda\approx 10^{-52}\mbox{\rm m}^{-2}\approx 10^{-35}\mbox{\rm s}^{-2}.
\end{equation}

In the paper of Behar and Carmeli a four-dimensional cosmological relativity 
theory that unifies space and velocity was proposed that
predicts the acceleration of the universe and hence it is equivalent to having
a positive value for $\Lambda$ in it. As is well known, in the traditional 
work of Friedmann when added to it a cosmological constant, the field 
equations obtained are highly complicated and no solutions have been obtained 
so far. Behar-Carmeli's theory, on the other hand, yields exact solutions and 
describes the universe as having a three-phase
evolution with a decelerating expansion followed by a constant and an
accelerating expansion, and it predicts that the universe is now in the 
latter phase. In the framework of this theory the zero-zero component of
Einstein's equations is written as
\begin{equation}
R^0_0-\frac{1}{2}\delta_0^0R=\kappa\rho_{eff}=\kappa\left(\rho-\rho_c\right),
\end{equation}
where $\rho_c=3/\kappa\tau^2\approx 3H^2/\kappa$ is the critical mass density. 
Comparing Eq. (32)
with the zero-zero component of Eq. (25), one obtains the expression for the
cosmological constant in the Behar-Carmeli theory, 
\begin{equation}
\Lambda=\kappa\rho_c=3/\tau^2\approx 3H^2.
\end{equation}

Assuming that Hubble's constant $H=70$km/s-Mpc, then $\Lambda=1.934\times 
10^{-35}$s$^{-2}$. This result is in good agreement with the recent supernovae
experimental results. The analyses presented in this paper for determining
the value of $\Lambda$ show that the same value for $\Lambda$ is obtained
here also, although the theory now is different. 

\end{document}